
\magnification 1200
\overfullrule 0 pt
\font\abs=cmr9
\font\ist=cmr8

\font\uit=cmu10

\def\CcC{{\hbox{\tenrm C\kern-.45em{\vrule height.65em width0.06em depth-.04em
\hskip.45em }}}}
\def\RrR{{\hbox{\tenrm I\kern-.17em{R}}}}
\def\HhH{{\hbox{\tenrm {I\kern-.18em{H}}\kern-.18em{I}}}}
\def\DdD{{\hbox{\tenrm {I\kern-.18em{D}}\kern-.36em {\vrule height.62em
width0.08em depth-.04em\hskip.36em}}}}
\def\ZzZ{{\hbox{\tenrm Z\kern-.31em{Z}}}}
\def\IiI{{\hbox{\tenrm I\kern-.19em{I}}}}
\def\NnN{{\hbox{\tenrm {I\kern-.18em{N}}\kern-.18em{I}}}}
\def\QqQ{{\hbox{\tenrm {{Q\kern-.54em{\vrule height.61em width0.05em
depth-.04em}\hskip.54em}\kern-.34em{\vrule height.59em width0.05em
depth-.04em}}
\hskip.34em}}}
\def\OoO{{\hbox{\tenrm {{O\kern-.54em{\vrule height.61em width0.05em
depth-.04em}\hskip.54em}\kern-.34em{\vrule height.59em width0.05em
depth-.04em}}
\hskip.34em}}}

\def\uq2{U_q({\uit su}(2))}

\def\fraz#1#2{{\strut\displaystyle #1\over\displaystyle #2}}

\def\part#1{\fraz{\partial}{\partial#1}}

\def\ii#1{\item{$\phantom{1}#1~$}}

\def\su2q{SU(2)_q}
\def\h1q{H(1)_q}

\def\nu{N_{1}}

\hsize= 15.5 truecm
\vsize= 22 truecm
\hoffset= 0.5 truecm
\voffset= 0 truecm

\null\vskip1.2truecm

\baselineskip= 13.75 pt
\footline={\hss\tenrm\folio\hss}
\centerline
{\bf  IDENTICAL PARTICLES AND PERMUTATION GROUP}
\bigskip
\centerline{
{\it    E.Celeghini ${}^1$, M.Rasetti ${}^2$ and G.Vitiello ${}^3$.}}
\bigskip
{\ist ${}^1$Dipartimento di Fisica - Universit\`a di Firenze and
INFN--Firenze, I 50125 Firenze, Italy}
\footnote{}{\hskip -.85truecm {\abs E--mail: CELEGHINI@FI.INFN.IT.
{}~~Manuscript of 8 pages. \hfill DFF 205/5/94 Firenze}}

{\ist ${}^2$Dipartimento di Fisica and Unit\`a INFM -- Politecnico di
Torino, I 10129 Torino, Italy}

{\ist ${}^3$Dipartimento di Fisica - Universit\`a di Salerno and INFN--Napoli,
I 84100 Salerno, Italy}

\bigskip
\bigskip
\bigskip

\noindent{\bf Abstract.} {\abs Second quantization is revisited and creation
and annihilation operators are shown to be related, on the same footing both
to the algebra ${\it h}(1)$, ${\underline {and}}$ to the superalgebra
${\it osp}(1|2)$ that are shown to be both compatible with Bose ${\underline
{and}}$ Fermi statistics.
The two algebras are completely equivalent in the one-mode sector but,
because of grading of ${\it osp}(1|2)$, differ in the many-particle case.
The same scheme is straightforwardly extended to the quantum case
 ${\it h}_q(1)$ and ${\it osp}_q(1|2)$.}

\bigskip
\bigskip
\bigskip

Claiming that a permutation of two particles has been performed requires
distinguishability of the particles themselves. An idealized operational
procedure to this effect would be for example the following: one first attaches
a label to each particle ({\it i.e.} a quantum number identifying its state) in
order to distinguish it from any other, then one interchanges the particles
and, finally, one looks once more at the labels, to make sure that the exchange
has been properly performed. However, one of the fundamental hypotheses of
quantum field theory is exactly that particles should be treated as identical
and indistinguishable. For this reason, the permutation group cannot be
related {\it ab initio} to second quantization but, as we shall show, is to be
introduced in the theory only at a second stage, by the physical request of
expressing the properties on $n$-particle states in terms of first quantization
observables.

This has deep consequences to the effect that the usual connection between
algebraic properties of second quantization operators and statistics of the
particles turns out to bear some arbitrariness. In order to prove this
statement, we shall build explicitly, in terms of $\underline{\rm
anticommuting}$ creation and annihilation operators, a theory where, by
imposing the symmetry or antisymmetry of the particle states, both bosons and
fermions can be simultaneously constructed. As briefly discussed at the end of
the paper, the construction presented should be considered as an example of a
much more general and far-reaching feature: since there is no necessary
connection between the observables over the Fock space and the particle
statistics, we are allowed not only to consider fermions and bosons as related
to the Weyl-Heisenberg algebra ${\it h}(1)$, but the scheme is extendable also
to more complex relation among observables ({\it e.g.} $\underline{\rm
quantum}$ algebras) and/or exotic statistics ({\it e.g.} anyons).  It should be
mentioned that the approach presented in this letter was inspired by the
property that most of the algebraic structures relevant to second quantization
physics are Hopf algebras$^{\, [1]}$: it is just the operation of coproduct,
characteristic of the latter but usually disregarded because primitive for Lie
algebras, recently brought to attention by studies of quantum algebras, which
stands at the basis of our conceptual construction.

More formally, let us begin by showing how creation and annihilation operators
can be related, on the same footing both to the algebra  ${\it h}(1)$, and to
the $\underline{\rm superalgebra}$ ${\it osp}(1|2)$. ${\it h}(1)$, is
customarily defined$^{\, [2]}$ to be generated by the four operators $( a,
a^{\dagger}, \IiI , N )$, with commutation relations
$$
[\, a\, ,\, a^\dagger \, ] = \IiI \, ,\quad\quad\, [\, N\, ,\, a \, ] = - a
\, , \quad\quad [\, N\, ,\, a^\dagger \,] = a^\dagger \, , \quad\quad\,
[\, \IiI \, ,\, \bullet \, ] = 0 \, .
\eqno{(1)}
$$
Upon characterizing the unitary representations ({\it i.e.} those for which
$N^\dagger = N$, $(a^{\dagger})^{\dagger} = a$) with lower bounded spectrum of
$N$ by the lowest eigenvalue $n_0$, one can write
$$\eqalign{
a^\dagger \, |k+n_0> &= \sqrt{k + 1} \, |k+n_0+1>\quad , \quad
a \, |k+n_0> = \sqrt{k} \, |k+n_0-1> \quad , \cr
N \, |k+n_0> &= (k+n_0) \, |k + n_0> \qquad, \qquad\qquad\qquad\qquad
\quad\quad\quad \; k \in \NnN \quad .\cr}
$$
The usual Fock space ${\cal F}$ is obtained for $n_0 = 0$, usually adopting
the relation $N \equiv a^\dagger a$ (one of the solutions of the equations
$[\, N\, ,\, a \, ] = - a \, , \, [\, N\, ,\, a^\dagger \,] = a^\dagger $):
$$\eqalign{
a^\dagger \, |n> &= \sqrt{n + 1} \, |n+1>  \quad , \quad a \, |n> = \sqrt{n}
\, |n-1> \quad , \cr
N \, |n> &= n \, |n>  \qquad, \qquad\qquad\qquad\quad\quad
\; n \in \NnN \quad .\cr}
 \eqno{(2)}           \,
$$
A related $\ZzZ_2$-{\sl graded} structure will be considered here, starting
from the set of three operators ${\cal S} \equiv ( a, a^{\dagger}, H )$
(notice that, as in (1), all operators are assumed to be unrelated), with $H$
{\sl even} and $a$ and $a^{\dagger}$ {\sl odd}. ${\cal S}$ is characterized by
the relations
$$
\{ \, a \, ,\, a^\dagger \, \} = 2 H \, ,\quad\quad [\, H\, ,\, a \, ] = - a
\, ,\quad\quad [\, H\, ,\, a^\dagger \, ] = a^\dagger \, , \eqno{(3)}
$$
and it is a {\sl subset}, $\underline{\rm not}$ a {\sl sub}-algebra, of the
$\ZzZ_2$-graded algebra ${\it osp}(1|2)^{\, [3]}$.
Completion of ${\cal S}$ to the whole ${\it osp}(1|2)$ in fact requires the
introduction of the
additional set ${\cal S}' \equiv ( J_- , J_+ )$ in the even sector,
$$
J_+ = {\scriptstyle{{1\over 2}}} \{ a^{\dagger}, a^{\dagger} \} \quad ,
\quad\quad\quad J_- = {\scriptstyle{{1\over 2}}} \{ a, a \} \quad ,\eqno{(4)}
$$
with commutation relations
$$
\eqalign{
&[ J_+ , a ] = -2 a^{\dagger} \quad , \quad\quad [ J_- , a^{\dagger} ] = 2 a
\quad , \qquad\qquad [ J_+ , a^{\dagger} ] = 0 = [ J_- , a ] \, ,
\cr
&[ J_+ , J_- ] = - 4 H \, , \quad\quad [ H , J_{\pm} ] = \pm 2 J_{\pm} \quad .
\cr }
\eqno{(5)}
$$
The bosonic sector ${\cal B} \equiv$  $( J_- , J_+ , {1\over 2} H )$ is, as
well known, isomorphic to $su(1,1)$, in the direct sum of the representations
$\kappa = {1\over 4} , {3\over 4}\; ^{[4]}$.

An explicit analysis shows that the set ${\cal S}$ with relations (3) is
sufficient to give rise to unitary representations of ${\it osp}(1|2)$ that
have the spectrum of $H$ bounded below,  and can be characterized by the lowest
eigenvalue of $H$, say $h_0$. Explicitly,
$$
\eqalign{
a^\dagger \; |h_0+2k+1> &\, =\quad \sqrt{2(k+1)} \, |h_0+2k+2> \quad , \cr
a^\dagger \; |h_0+2k>\quad\;\; &\, =\, \sqrt{2(k+h_0)} \; |h_0+2k+1>
\quad , \cr
a \; |h_0+2k+1> &\, =\, \sqrt{2(k+h_0)} \; |h_0+2k> \quad , \cr
a \; |h_0+2k>\quad\;\; &\, =\quad\; \sqrt{2k} \quad\, |h_0+2k-1>
\quad , \cr
H \, |h_0+k>\quad\quad\, &\, =\quad (h_0+k) \;\; |h_0+k> \quad ,
\quad\quad\quad\quad\quad
\; k \in \NnN \quad ,\cr}
\eqno{(6)}
$$
where the partition of states in two classes exhibits the existence of
supersymmetric doublets.

The main point of our derivation is the fact that eqs.(6), with $h_0 =
{1\over 2}$, read
$$
\eqalign{
a^\dagger \,|h> &\, =\, \sqrt{h+{\scriptstyle{1\over 2}}} \, |h+1> \quad ,\cr
a \, |h> &\, =\, \sqrt{h-{\scriptstyle{1\over 2}}} \, |h-1> \quad , \cr
H \, |h> &\, =\quad h \quad |h> \quad ,\quad\qquad\qquad\quad
\quad\quad\quad h \in \NnN + {\scriptstyle{{1\over 2}}} \quad , \cr}
\eqno{(7)}
$$
namely they coincide with eqs.(2), provided the identification $h = n + {1
\over 2}$ is implemented. This means that the closed subset ${\cal S}$ of ${\it
osp}(1|2)$ defined by (3) shares the representation (2) -- which can by
induction be extended to the whole ${\it osp}(1|2)$ -- with the Weyl-Heisenberg
algebra ${\it h}(1)$. In other words, the Fock space ${\cal F}$ provides a
faithful representation for both ${\it h}(1)$ (for $n_0 = 0$) and ${\it
osp}(1|2)$ (for $h_0 = {1\over 2}$).

Second quantization is based, essentially, on the relations (2). We
suggest that creation and annihilation operators may therefore be, with equal
rights,
be interpreted as belonging either to ${\it osp}(1|2)$ or to ${\it h}(1)$.
The key point in our argument is the following: as far as one considers the
algebra as generated by the defining commutation relations only, any physical
interpretation is contained in eqs. (2), and it turns out to be essentially
irrelevant whether one selects ${\it osp}(1|2)$ or ${\it h}(1)$.  However, when
one deals with many-particle states, the two schemes lead to self-consistent
yet mutually unequivalent descriptions.  The reason why this may happen, is
that ${\it h}(1)$ and ${\it osp}(1|2)$ are Hopf algebras (more precisely, ${\it
osp}(1|2)$ is a {\sl super} Hopf algebra). Any Hopf algebra, say ${\cal A}$,
has, among its defining operations, the coproduct $\Delta :{\cal A} \to {\cal
A} \otimes {\cal A}$ (in fact, in the representation considered here, this is
necessary, in that it implies that the action of the algebra is well defined on
${\cal F} \otimes {\cal F}$ and, by induction, on  ${\cal F}^{\otimes n}$). For
both ${\it h}(1)$ and ${\it osp}(1|2)$ $\Delta$ is, of course, primitive.

In ${\it h}(1)$ one has:
$$
\eqalign{
&\Delta(a) = a \otimes {\bf 1} + {\bf 1} \otimes a \equiv a_1 + a_2 \quad ,
\qquad \Delta(a^{\dagger}) = a^{\dagger} \otimes {\bf 1} + {\bf 1} \otimes
a^{\dagger} \equiv a_1^{\dagger} + a_2^{\dagger} \; , \cr
&\Delta(N) = N \otimes {\bf 1} + {\bf 1} \otimes N \equiv N_1 + N_2 \; ,
\quad \Delta(\IiI) = \IiI \otimes {\bf 1} + {\bf 1} \otimes \IiI \quad ,
\cr}\eqno{(8)}
$$
(notice that $\IiI$ is a central operator with eigenvalue 1 and $\underline{
\rm not}$ the identity ${\bf 1}$ of the universal enveloping algebra ($UEA$)).

The coalgebra for the superalgebra ${\it osp}(1|2)$ looks quite similar:
$$
\eqalign{
\Delta(a)\; &= a \otimes {\bf 1} + {\bf 1} \otimes a \equiv a_1 + a_2 \,
,\qquad
\qquad \Delta(a^{\dagger}) = a^{\dagger} \otimes {\bf 1} + {\bf 1} \otimes
a^{\dagger} \equiv a_1^{\dagger} + a_2^{\dagger} \, , \cr
\Delta(H) &= H \otimes {\bf 1} + {\bf 1} \otimes H \equiv H_1 + H_2 \, ,\cr
} \eqno{(9)}
$$
however, since $a$ and $a^{\dagger}$ are odd, whereas $H$ is even, we have
for $c,d, e, f \in {\it osp}(1|2)$, the multiplication law on ${\cal F} \otimes
{\cal F}$
$$
(c \otimes d) (e \otimes f)\, = \, {(-)}^{p(d) p(e)}\; c e \otimes d f \quad ,
\eqno{(10)}
$$
where $p(d)$ and $p(e) \in \ZzZ_2$ are the degrees ({\it i.e.} parities) of
$d$ and $e$ respectively.
On ${\cal F}^{\otimes n}$ composition rules are therefore quite different.
Let us denote
$$
\eqalign{
a_j \, &\equiv \, {\bf 1} \otimes {\bf 1} \otimes \dots \otimes {\bf 1}
\otimes a \otimes {\bf 1} \otimes \dots \otimes {\bf 1} \quad , \cr
{a_j}^\dagger \, &\equiv \, {\bf 1} \otimes {\bf 1} \otimes \dots \otimes
{\bf 1} \otimes a^\dagger \otimes {\bf 1} \otimes \dots \otimes {\bf 1}
\quad , \cr
H_j \, &\equiv \, {\bf 1} \otimes {\bf 1} \otimes \dots \otimes {\bf 1}
\otimes H \otimes {\bf 1} \otimes \dots \otimes {\bf 1} \quad , \cr
N_j \, &\equiv \, {\bf 1} \otimes {\bf 1} \otimes \dots \otimes {\bf 1}
\otimes N \otimes {\bf 1} \otimes \dots \otimes {\bf 1} \quad , \cr}
$$
where the multiple $\otimes$-products have $n$ factors, in each of which the
only element different from the $UEA$ identity ${\bf 1}$ is in the  $j-th$
position.
One has, for $( a , a^\dagger , N , \IiI )$ in ${\it h}(1)$ the
customary relations $[ a_i , a_j ] = 0$, $[ a_i , a_j^\dagger ] =
\delta_{i \, j} \IiI$, $[ N_i , a_j ] = - a_i \delta_{i\, j}$ (plus their
hermitian conjugates), while for $( a , a^\dagger , H )$ in ${\it osp}(1|2)$,
the (graded) commutation relations are
$$
\eqalign{
\{ a_i , {a_j}^{\dagger} \} \, =\, 2 H_i\, \delta_{ij} , \quad ,\quad [ H_i ,
a_j ] \, &=\, - a_i \, \delta_{ij} , \quad ,\quad [ H_i , {a_j}^{\dagger} ] \,
=\; {a_i}^{\dagger} \, \delta_{ij} \quad , \cr
\{ {a_i}^{\dagger}, {a_j}^{\dagger} \} =\; \delta_{ij} J_+ \quad &,\quad
\{ a_i , a_j \} \, =\, \delta_{ij} J_- \quad , \cr}
$$
and, of course, $[ a_j , a_j ] \, =\, 0$.

On the Fock basis of ${\cal F}^{\otimes n}$ adoption of ${\it osp}(1|2)$ leads
to
$$
\eqalign{
{a_j}^{\dagger} &|n_1,..,n_{j-1},n_j,..,n_n> \, =\,
(-1)^{s_j}  \sqrt{n_j+1}\, |n_1,..,n_{j-1},n_j+1,..,n_n> \quad , \cr
a_j~ &|n_1,..,n_{j-1},n_j,..,n_n> \, =\,
(-1)^{s_j} \sqrt{n_j}\, |n_1,..,n_{j-1},n_j-1,..,n_n> \quad ,\cr
N_j~ &|n_1,..,n_{j-1},n_j,..,n_n> \;=\quad
n_j\quad |n_1,..,n_{j-1},n_j,..,n_n> \quad , \cr}
\eqno{(11)}
$$
where the phases turn out to be exactly those customarily used in textbooks
for $\underline{\rm fermions}^{\; [5]}$  $\displaystyle{\Bigl (s_j \equiv
\sum_{k=1}^{j-1} n_k\Bigr )}$.

It is worth pointing out that eqs.(11) differ from the usual (bosonic) ones
only in the choice of phases and, on ${\cal F}^{\otimes n}$, imply $[ a_i ,
{a_i}^{\dagger} ] = \IiI$, consistently with the standard formulation, which in
turn gives $\{ a_i , {a_i}^{\dagger} \} = 2 H_i \; (\equiv 2 N_i+1)$.
Nevertheless the subtle and important implication here is that, in order to
determine the phases of the basis vectors, an order must be imposed {\it a
priori} in the set of indices $j$'s, in such a way that
$$
|n_1,..,n_j,..,n_n> \, \equiv {1\over{\sqrt{\prod_{j=1}^n n_j !}}} \;
({a_1}^{\dagger})^{n_1} \dots ({a_j}^{\dagger})^{n_j} \dots
({a_n}^{\dagger})^{n_n} \, |0> \quad ,
$$
contrary to the standard bosonic theory, where the creation operators commute
and can be applied in any order.  Of course, the usual properties of the Fock
space, such as completeness
$$
\sum_{\{ n\} } |n_1, n_2,\dots > <n_1, n_2,\dots | = \IiI \quad ,
$$
and the projection operators on the one- and two-particles states
$$
{\cal P}_1 \equiv \sum_i |1_i><1_i| \quad ,\quad {\cal P}_2 \equiv
\sum_{i<j} |1_i, 1_j> <1_i, 1_j| + \sum_i |2_i> <2_i| \quad ,
\eqno{(12)}
$$
where
$$
\eqalign{
|1_i>\quad\; &\equiv\, |n_1=0, n_2=0,\dots , n_{i-1}=0 , n_i=1, n_{i+1}=0 ,
\dots , n_n=0> \quad , \cr
|2_i>\quad\; &\equiv\, |n_1=0, n_2=0,\dots , n_{i-1}=0 , n_i=2, n_{i+1}=0 ,
\dots , n_n=0> \quad , \cr
|1_i, 1_j> &\equiv\, |n_1=0 , n_2=0,\dots, n_i=1 , \dots , n_j=1 , \dots n_n=0>
\quad , \cr}
$$
do not depend on such phases, and persist. ${\it osp}(1|2)$ can, in such a
way, be utilized to construct $n$-particle states, leading to a scheme
non-equivalent to that derivable from ${\it h}(1)$, because of the grading of
{\it odd} operators.

This possibility of using anticommuting operators, without restriction on the
occupation numbers $n_j$'s, also casts a new light on the question of how one
should introduce statistics into play.

As stressed in the introduction, second quantization is essentially
uncorrelated with statistics$^{\, [6]}$, which is required as a necessary set
of rules to represent isomorphically $n$-particle states in the state space
given by the $n$-fold tensorization of the single-particle Hilbert space,
proper of first quantization.  The relevant point in the analysis of this
problem performed by Pauli in [6] is that the symmetry with respect to
permutation of two particles does not depend on the prescription to build
${\cal F}$ from the vacuum ({\it i.e.} on the commutation or anticommutation
relations of the $a_i$'s and ${a_j}^{\dagger}$'s), but it must be imposed as an
external constraint aiming to guarantee a correct implementation of the above
isomorphism.

In such a perspective, let us consider what happens with two $\underline{
\rm bosons}$. Independently on whether the algebra is graded or not, one has to
consider for such a system a symmetric Hilbert space, namely a generic state
vector must be symmetric with respect to the exchange of the two particles
(this being the feature which qualifies them as boson)
$$
{|x_1, x_2 >}_{B} \, \equiv \, {\scriptstyle{{1\over \sqrt{2}}}} \left (
|x_1>\, |x_2> + |x_2>\,  |x_1> \right ) \quad ,
$$
and, by (12),
$$
{|x_1, x_2>}_{B} \, = \, \sum_{i<j} {|1_i, 1_j><1_i, 1_j|x_1, x_2>}_{B} +
\sum_i {|2_i><2_i|x_1, x_2>}_{B} \quad .
$$
Independently on how the states  $|1_i, 1_j>$ and $|2_i>$ are constructed from
the vacuum, the symmetry is here automatically implemented, in that
$$
\eqalign{
{<1_i, 1_j|x_1, x_2>}_{B} &\,=\,  {\scriptstyle{{1\over \sqrt{2}}}}
<1_i, 1_j|\left ( |x_1, x_2> + |x_2, x_1>\right )\cr
&\, =\, {\scriptstyle{{1\over \sqrt{2}}}} \left ( <1_i|x_1><1_j| x_2> +
<1_i|x_2><1_j| x_1>\right )\quad ,\cr
{<2_i|x_1, x_2>}_{B} &\, =\,  {\scriptstyle{{1\over \sqrt{2}}}}
<2_i|\left ( |x_1, x_2> + |x_2, x_1>\right ) =\, <1_i|x_1><1_i| x_2>
\quad ,\cr} \eqno{(13)}
$$
manifestly have the required invariance with respect to interchange of the
two particles.

The feature that statistics has no connection with the algebra is further
proved by the fact that ${\it osp}(1|2)$, as well as ${\it h}(1)$, work equally
well with $\underline{ \rm fermions}$. For two fermions we must consider an
antisymmetric Hilbert state-space
$$
{|x_1, x_2 >}_{F} \equiv {\scriptstyle{{1\over \sqrt{2}}}} \left ( |x_1>
\, |x_2> - |x_2>\, |x_1>\right )\quad ,
$$
and, once more independently on the algebra considered, the antisymmetry in
the exchange of the two fermions is guaranteed, as well as the Pauli exclusion
principle:
$$
\eqalign{
{<1_i, 1_j|x_1, x_2>}_{F} &\, =\, {\scriptstyle{{1\over \sqrt{2}}}}
<1_i, 1_j|\left (|x_1, x_2> - |x_2, x_1>\right ) \cr
&\, =\, {\scriptstyle{{1\over \sqrt{2}}}} \left ( <1_i|x_1><1_j| x_2> -
<1_i|x_2> <1_j| x_1>\right )\quad , \cr
{<2_i|x_1, x_2>}_{F} &\, =\, {\scriptstyle{{1\over \sqrt{2}}}} <2_i|\left (
|x_1, x_2> - |x_2, x_1>\right ) \cr
&\, =\,  {\scriptstyle{{1\over \sqrt{2}}}} \left (<1_i|x_1><1_i| x_2> -
<1_i|x_2><1_i| x_1>\right ) = 0 \quad .\cr}
\eqno{(14)}
$$

\noindent (13) and (14) clearly demonstrate the novel possibility -- besides
the
customary scheme$^{\, [4]}$ -- of constructing bosons with graded operators
or fermions with even operators.

It should be stressed that our arguments in this paper are quite different from
other procedures whereby $''${\sl ad hoc}$''$ constraints are introduced on the
variables in order to generate the statistics. An instance of such different
approaches are non-linear transformations, along the lines proposed by
Gutzwiller's projection operator method$^{[7]}$. In the fermionic case, a
suggestive example is provided by the new creation and annihilation operators
defined by
$$
c_j = a_j {\cal Q}_j \quad , \quad {\cal Q}_j \equiv {\scriptstyle{{1\over
2}}} {{\left ( {\bf 1} - {\rm e}^{i \pi N_j} \right )}\over{\sqrt{
N_j + {1\over 2} \left ( {\bf 1} + {\rm e}^{i \pi N_j} \right ) }}} = {\cal
Q}_j^{\dagger} \quad , \quad c_j^{\dagger} = {\cal Q}_j a_j^{\dagger} \quad .
\eqno{(15)}
$$

It is straightforward to check that -- because of (2) -- (15) leads to both the
Pauli exclusion principle, $c_j^2 = 0 = ( c_j^\dagger )^2$, and the customary
fermionic anticommutation relations $\{ c_i, {c_j}^{\dagger} \} = \delta_{i\,
j} \IiI$, on the subsector of ${\cal F}^{\otimes n}$ consisting of paired
superdoublets $\{ |\dots , 2 n_j , \dots > , |\dots , 2 n_j + 1 , \dots > \}$.
The usual fermions can therefore be recovered by restriction to $n_j = 0$.
Of course, in the above procedure no reference or use has been done of grading.

Adoption of a graded algebra has also deep bearings on the structures one can
induce in the $UEA$. For example, it is usually assumed that the algebra
$su(1,1)$ can be constructed in the $UEA$ of ${\it h} (1)$.  This is not true:
as stressed before $su(1,1)$ can be easily obtained from eqs.(3) as the bosonic
sector of the superalgebra ${\it osp}(1|2)$, while one has to use $n_0 = 0$
({\it i.e.} one needs to impose indirectly the ${\it osp}(1|2)$ properties
also) to obtain the same result from eqs.(1). Moreover, the grading property
(10) plays an essential r\^ole in obtaining the coalgebra (of course primitive)
of $su(1,1)$ from the one of  ${\cal S}$, while it is impossible to obtain the
same result from the coalgebra of ${\it h}(1)$.

Indeed, from (5), (8) and (9), we have, for instance, in  ${\it h} (1)$
$\Delta(J_-) = a^2 \otimes {\bf 1} + {\bf 1} \otimes a^2 +  2\, a \otimes a$
whereas in  ${\it osp}(1|2)$, because of (11), $\Delta(J_-) = a^2 \otimes {\bf
1} + {\bf 1} \otimes a^2 \equiv J_- \otimes {\bf 1} + {\bf 1} \otimes J_-$, as
it should, because $J_-$ is primitive.

This shows that  $su(1,1)$ is contained as a full Hopf algebra in the universal
envelope of ${\cal S}$, while only in the common representation (2), the
$su(1,1)$ algebra can be considered as realized in the $UEA$ of ${\it h}(1)$
(of course, all these can be extended to  $su(2)$ by analytical continuation
from $su(1,1)$).

We have now to recall that both ${\it h}(1)^{\, [8]}$ and ${\it osp} (1|2)^{\,
[9]}$ have  $\underline{\rm quantum}$ deformations and the whole discussion
could be easily extended to them. Actually, it should be kept in mind that, in
the scheme proposed, no relation links the algebraic features of the creation
and annihilation operators to the symmetry of the states. It is, indeed,
possible to study systems of particles, both fermions or bosons, by means of
either ${\it h}_q(1)$ or ${\it osp}_q(1|2)$. The Fock space remain the same,
while differences appear principally in the relations of composed observables
with the single-particle ones.

We finally conjecture that in the present approach, there is room for
considering objects with more complex symmetry such as anyons.

\bigskip
\bigskip
\line{{\bf Acknowledgements}\hfill}
\noindent{\abs The authors gratefully acknowledge fruitful discussions with
F. Iachello.}
\bigskip
\bigskip
\bigskip
\bigskip

\line{{\bf References.}\hfill}
\bigskip

\ii {[1]} J. Fuchs, {\sl Affine Lie Algebras and Quantum Groups}, Cambridge
University Press, Cambridge, 1992

\ii {[2]} W. Miller, {\sl Lie  Theory and Special Functions}, Academic
Press, New York, NY, 1968

\ii {[3]} V.G. Kac, Comm. Math. Phys. {\bf 53}, 31 (1977)

\ii {} P. Ramond, Physica {\bf D 15}, 25 (1985)

\ii {[4]} A. Perelomov, {\sl Generalized Coherent States and Their
Applications}, Springer Verlag, Berlin, 1986

\ii {[5]} S.S.Schweber, {\sl An Introduction to Relativistic Quantum Field
Theory}, Row Peterson, Evanston ILL, 1961

\ii {[6]} W. Pauli, {\sl Handbuch der Physik} {\bf 24} 1.Teil, 83 (1933)
(english translation {\sl General Principles of Quantum
Mechanics}, Springer, Berlin, 1980)

\ii {[7]} M. Gutzwiller, Phys. Rev. {\bf A 137}, 1726 (1965)

\ii {[8]} E. Celeghini, R. Giachetti, E. Sorace, and M. Tarlini, J. Math.
Phys. {\bf 31}, 2548 (1990), and {\bf 32}, 1155 (1991)

\ii {[9]} P.P. Kulish, and N.Yu. Reshetikin, Lett. Math. Phys. {\bf 18},
143 (1989)

\ii {} E.Celeghini, T.D.Palev and M.Tarlini, Mod. Phys. Lett {\bf B5},
187 (1991)

\vfill\eject
\bye